\documentclass[11pt]{article}

\usepackage{amsfonts}
\usepackage{amssymb}
\usepackage{amsmath}
\usepackage{amsthm}
\usepackage{amsfonts}
\usepackage{multicol}
\usepackage{amscd}
\usepackage{textcomp}
\usepackage{multirow}
\usepackage[english, activeacute]{babel}

\def\be{\begin{equation}}
\def\ee{\end{equation}}
\def\bea{\begin{eqnarray}}
\def\eea{\end{eqnarray}}

\def\pois#1#2{\left\{ {#1},{#2} \right\}}         
         
\def\conm#1#2{\left [ {#1},{#2} \right ]}

\def\b{\mathfrak{r}_3(1)}

\def\We{e}
\def\Ge{\mathfrak{e}}

\def\1{\'{\i}}

\def\>#1{{\mathbf#1}}

\def\pa{\varphi}            
\def\ph{h}

\begin{document}

\thispagestyle{empty}


\ 
\vspace{0.5cm}

\begin{center}

{\Large{\sc{Non-coboundary Poisson-Lie structures on the book group
}} }

\end{center}

\medskip

\begin{center} \'Angel Ballesteros, Alfonso Blasco and Fabio Musso
\end{center}

\begin{center} {\it {Departamento de F\1sica,  Universidad de Burgos, 
09001 Burgos, Spain}}

e-mail: angelb@ubu.es, ablasco@ubu.es, fmusso@ubu.es
\end{center}

  \medskip

\begin{abstract} 
\noindent
All possible Poisson-Lie (PL) structures on the 3D real Lie group generated by a dilation and two commuting translations are obtained.  Their classification is fully performed by relating these PL groups with the corresponding Lie bialgebra structures on the corresponding `book' Lie algebra.  By construction, all these Poisson structures are quadratic Poisson-Hopf algebras for which the group multiplication is a Poisson map. In contrast to the case of simple Lie groups, it turns out that most of the PL structures on the book group are non-coboundary ones. Moreover, from the viewpoint of Poisson dynamics,  the most interesting PL book structures are just some of these non-coboundaries, which are explicitly analysed. In particular,  we show that  the two different $q$-deformed Poisson versions of the $sl(2,R)$ algebra appear as two distinguished cases in this classification, as well as the quadratic Poisson structure that underlies the integrability of a large class of 3D Lotka-Volterra equations. Finally, the quantization problem for these PL groups is sketched.

 \end{abstract}

\bigskip\bigskip\bigskip\bigskip





\section{Introduction}

Poisson-Lie (PL) groups are Poisson structures on a Lie group $G$ for which the group multiplication is a Poisson map. They were introduced by Drinfel'd~\cite{DrPL, DrICM} in connection with quantum groups, which are just Hopf algebra quantizations of PL structures (see~\cite{DrPL}-\cite{Bidegain} and references therein). Therefore, PL groups are Poisson-Hopf algebras endowed with a coproduct map $\Delta$ 
\be
\Delta: C^\infty(G)\rightarrow C^\infty(G)\otimes  C^\infty(G)
\ee
which is directly induced from the group multiplication. In such Hopf algebraic setting  the associativity of the group law for $G$ is translated into the coassociativity property for $\Delta$
\be
(\Delta\otimes id)\circ \Delta = (id\otimes \Delta)\circ \Delta
\ee
and the Poisson map condition for the group multiplication is written as 
the homomorphism condition for the coproduct map
\be
\pois{\Delta (a)}{\Delta (b)}=\Delta(\pois{a}{b}) \qquad
\forall\,a,b\,\in C^\infty(G).
\ee

The main classification result for PL groups (and, essentially, for quantum groups) states that Poisson-Lie structures on a (connected and simply connected) Lie group $G$ are in one-to-one correspondence with the Lie bialgebra structures on the Lie algebra $g=\mbox{Lie}(G)$~\cite{DrPL}.
We recall that a Lie bialgebra $(g,\delta)$ is Lie algebra $g$ endowed with a
skewsymmetric cocommutator map
$\delta:{g}\to {g}\otimes {g}$ such that

i) The map $\delta$ is a 1--cocycle, i.e.,
\be
\delta([X,Y])=[\delta(X),\, 1\otimes Y+ Y\otimes 1] + 
[1\otimes X+ X\otimes 1,\, \delta(Y)] \qquad \forall \,X,Y\in
{g}.
\ee

ii) The dual map $\delta^\ast:{g}^\ast\otimes {g}^\ast \to
{g}^\ast$ is a Lie bracket on ${g}^\ast$.

Given a PL structure on $G$, its associated `tangent' Lie bialgebra structure $(g,\delta)$ on $g=\mbox{Lie}(G)$ can be straightforwardly obtained. Firstly, the Poisson bracket has to be written in terms of the  local coordinates dual to the Lie algebra generators. Afterwards, the linearization of the PL bracket written in local coordinates defines the dual of the cocommutator map
\be
\delta^\ast:{g}^\ast\otimes {g}^\ast \to
{g}^\ast
\ee
from which $\delta$ can be recovered.

A Lie bialgebra $(g,\delta)$ is called a {\em coboundary} one when there exists a skew\-symmetric element $r$ of  ${g}\otimes {g}$ (the classical $r$-matrix) such that  the cocommutator $\delta$ is given by 
\be
\delta(X)=[1\otimes X + X \otimes 1,\,  r]\qquad 
X\in {g}.
\label{cocor}
\ee
We stress that in order to ensure that \eqref{cocor} fulfills properties i) and ii), the Schouten bracket $[[r,r]]$ has to be a constant solution of the modified classical Yang--Baxter equation (mCYBE)
\be
[X\otimes 1\otimes 1 + 1\otimes X\otimes 1 +
1\otimes 1\otimes X,[[r,r]]\, ]=0 \qquad X\in {g}.
\label{mcybe}
\ee
Therefore, the classification (modulo equivalence under automorphisms) of the constant skew\-symmetric solutions of the mCYBE on a given Lie algebra $g$ provide the classification of coboundary PL structures on the (connected and simply connected) group G. For semisimple Lie algebras all Lie bialgebra structures are coboundaries, and the PL classification problem can be fully solved through the classification of classical $r$-matrices.
Moreover, this correspondence is fully constructive since the explicit PL structure associated to a given $r$ is provided by the Sklyanin bracket
\be
\pois{f}{g}= r^{\alpha\beta}\left(X_{\alpha}^L f \,X_{\beta}^L g - 
X_{\alpha}^R f \,X_{\beta}^R g \right) 
\qquad
f,g\in C^\infty(G)
\label{sb}
\ee
where $X_{\alpha}^L$ and $X_{\alpha}^R$ are, respectively, left and right invariant vector fields on $G$.

But if the group $G$ is not semisimple then {\em non-coboundary} Lie bialgebras (and, therefore, non-coboundary PL groups) can appear. Indeed, a glimpse on the paper~\cite{gomez} (in which the Lie bialgebra structures for all 3D real Lie algebras are fully classified) makes evident that non-coboundaries are the {\em dominant} structures for 3D non-semisimple PL groups. We stress that in such non-coboundary cases no Sklyanin bracket is available and the corresponding PL structures have not been -in most of the cases- constructed. Despite of this fact, the previous literature on PL groups is mainly concentrated on the coboundary cases, probably due to their universality for semisimple groups and also to the fact that their associated quantum groups can be quantized by using a quantum $R$-matrix of the form $R=1+\hbar\,r + o[\hbar^2]$ (the so-called FRT approach~\cite{FRT}).

The aim of this paper is to get a deeper insight into non-coboundary Poisson-Lie groups by studying the classification and explicit construction of all PL group structures on the 3D real solvable  `book group'~\cite{LuW,CMP94}. We will show that, firstly, all non-coboundary PL structures can be explicitly obtained; secondly, that these PL groups present a wide range of applications and, finally, that the Hopf algebra quantization for non-coboundary structures is also feasible.

In the next section the definition and notations concerning the book Lie group and algebra are presented. In section 3 all possible PL structures on the book group are obtained by direct computation as the set of all possible quadratic Poisson structures that are compatible with the coproduct induced by the book group multiplication law. This result is put into exact correspondence with the equivalence classes of Lie bialgebra structures on the book Lie algebra given in~\cite{gomez}, that contains 9 classes of PL structures from which 7 of them are non-coboundaries. The next sections are devoted to the specific analysis of three distinguised non-coboundary PL book groups. In section 4 we show that the first of them underlies the integrability of a wide class of 3D Lotka-Volterra (LV) equations and, from the perspective proposed in~\cite{BBMLV}, the generic multiparametric PL book group obtained in section 3 provides integrable perturbations of such set of LV equations. Also, the Hopf algebra quantization of this LV structure is obtained, leading to a quantum euclidean algebra. In section 4, by performing a suitable change of local coordinates on the book group that includes a new `quantum' deformation parameter, the Poisson version of the standard $q$-deformation of $sl(2)$ is obtained as one of the equivalence classes of PL structures for specific values of the parameters. We mention that this result was already obtained in~\cite{Marmo}, but here we put it into a global classification perspective that stresses the role of the Lie bialgebra parameters. Similarly, in section 5 we work out in detail the Poisson version of the non-standard $q$-deformation of $sl(2)$ as a third distinguished non-coboundary PL book group, and we recall the applications of this Poisson structure in order to construct integrable models on $N$-dimensional curved spaces. Section 6 contains a summary of the properties of the remaining six classes of PL book groups, including the coboundary ones, together with several comments concerning their quantization. The paper is closed by a final section in which several remarks and future research objectives are presented.


\section{The book Lie algebra and group}

The `book' Lie algebra $\b$   is defined by the Lie brackets
\be
[e_1,e_3]=e_1 \qquad [e_2,e_3]=e_2 \qquad [e_1,e_2]=0.
\label{bookcr}
\ee
Note that although the `book' name is taken from~\cite{LuW,CMP94} and the notation $\b$ from~\cite{gomez}, this algebra is the Bianchi V Lie algebra, the $A_{3,3}$ one in~\cite{pavel} and $\mathfrak{an}(2)$ in~\cite{MSjmp}.
Therefore \eqref{bookcr} defines a solvable 3D real Lie algebra where $e_3$ can be interpreted as the generator of dilations while both $e_1$ and $e_2$ are flat Euclidean translations.
Its adjoint representation $\rho$ is
\be
\rho(e_1)=\left( 
\begin{array}{ccc}
0 & 0 & 1\\
0 & 0 & 0\\
0 & 0 & 0
\end{array}
\right)  \quad 
\rho(e_2)=\left( 
\begin{array}{ccc}
0 & 0 & 0\\
0 & 0 & 1\\
0 & 0 & 0
\end{array}
\right) \quad
\rho(e_3)=\left( 
\begin{array}{ccc}
-1 & 0 & 0\\
0 & -1 & 0\\
0 & 0 & 0
\end{array}
\right)
\label{repbook}
\ee
and the generic Lie group element $M$ with `local coordinates'  given by $(y,z,x)$ reads
\be
M=\exp \left(y \rho(e_1) \right) \exp \left(z \rho(e_2) \right) \exp \left(x \rho(e_3) \right)=
\left( 
\begin{array}{ccc}
\exp(-x) & 0 & y\\
0 & \exp(-x) & z\\
0 & 0 & 1
\end{array}
\right).
\ee

If we consider the new variables 
\be
X=\exp(-x)
\qquad
Y=y
\qquad
Z=z
\label{localchange}
\ee
we get
\be
M=\left(
\begin{array}{ccc}
 X & 0 & Y \\
 0 & X & Z\\
 0 & 0 & 1
\end{array} 
\right)
\label{group}
\ee
and the multiplication of two group elements would be
\be
M_1\cdot M_2=\left(
\begin{array}{ccc}
 X_1\,X_2 & 0 & X_1\,Y_2 + Y_1\\
 0 & X_1\,X_2 & X_1\,Z_2 +Z_1 \\
 0 & 0 & 1
\end{array} 
\right).
\ee
The coproduct $\Delta: C^\infty(G)\rightarrow C^\infty(G)\otimes  C^\infty(G)$ of the coordinate functions on the group is the pullback of the group multiplication, therefore
\begin{equation}
\label{bookcopr}
\begin{array}{lcl}
\Delta(X)(M_1,M_2) \equiv X(M_1 \cdot M_2)=X_1 \, X_2 \qquad & \Rightarrow  \qquad & \Delta(X)=X \otimes X,  \\
\Delta(Y)(M_1,M_2) \equiv Y(M_1 \cdot M_2)=X_1 \, Y_2+ Y_1 \qquad & \Rightarrow  \qquad & \Delta(Y)=X \otimes Y + Y \otimes 1,\\
\Delta(Z)(M_1,M_2) \equiv Z(M_1 \cdot M_2)=X_1 \, Z_2+ Z_1 \qquad & \Rightarrow  \qquad & \Delta(Z)=X \otimes Z + Z \otimes 1. 
\end{array}
\end{equation}
Equivalently, in `local' coordinates $(z,y,x)$ the coproduct map would be
\bea
\Delta(x)&=& 1 \otimes x + x \otimes 1\nonumber \\
\Delta(y)&=&e^{-x} \otimes y + y \otimes 1
\label{bookcoprlocal}\\
\Delta(z)&=&e^{-x} \otimes z + z \otimes 1 
\notag
\eea
and since $\Delta(e^{-x})=e^{-x}\otimes e^{-x}=e^{-\Delta(x)}$, this implies that $\Delta(x)= 1 \otimes x + x \otimes 1$.

Within this algebraic perspective, given a Poisson-Lie structure $\cal{P}$ on $C^\infty(G)$, the condition that the group multiplication is a Poisson map for $\cal{P}$ is rewritten as
\be
\pois{\Delta (a)}{\Delta (b)}=\Delta(\pois{a}{b}) \qquad
\forall\,a,b\,\in C^\infty(G)
\ee
which means that $(C^\infty(G),{\cal P},\Delta )$ is a Poisson coalgebra (a Poisson-Hopf algebra if the counit and antipode maps are included~\cite{CP,majid}).


\section{Classification of PL structures on the book group}

The following result can be proven through a long but straightforward computation:

\noindent {\bf Proposition}. The most generic quadratic Poisson structure in $\{X,Y,Z,1\}$ for which the comultiplication $\Delta$ \eqref{bookcopr} is a Poisson map is given by the brackets
 \begin{eqnarray}
\{X,Y\}&=&a X^2-b X Y-2 c X Z-a X \notag\\
 \{X,Z\}&=&d X^2+2 e X Y+b X Z-d X
 \label{genericPL}\\
\{Y,Z\}&=& -f X^2+e Y^2+b Y Z-d Y+c Z^2+a Z+f.\notag
\end{eqnarray}
We will call this six-parametric Poisson bivector as ${\cal P}{[a,b,c,d,e,f]}$. Moreover, the generic Casimir function for \eqref{genericPL} is found to be
\be
\mathcal{C}=
 \dfrac{
f(1+X^{2})+(X-1)(dY-aZ)+e Y^{2}+(b Y+c Z)Z}
{X}
\label{casbook}
\ee
\medskip
Consequently, we can say that $({\cal P}{[a,b,c,d,e,f]},\Delta)$ is  the most generic PL structure on $C^\infty(G)$.

As we shall see in the sequel, it is also convenient to write ${\cal P}{[a,b,c,d,e,f]}$ in the local coordinates $(z,y,x)$ given by \eqref{localchange}. We get
\bea
&& \pois{x}{y}=  a (1-e^{-x})+ by+2c z
\cr
&& \pois{x}{z}=  d (1-e^{-x}) - 2 e y- b z
\label{pllocal}\\
&& \pois{y}{z}= f (1-e^{-2x})+e y^2+b y z-d y+c z^2+a z
\nonumber
\eea
and the Casimir function is now
\be 
\mathcal{C}= e^{x}\left[
f(1+e^{-2x})+d(-1+e^{-x})y+e y^{2}+a z(1-e^{-x})+z(b y+c z)
\right].
\ee
In these local coordinates the linearization ${\cal P}_0$ of \eqref{pllocal} is straightforwardly obtained as
\bea
&& \pois{x}{y}_0=  a x + b y+ 2 c z
\cr
&& \pois{x}{z}_0=  d x - 2 e y - b z
\label{p0}\\
&& \pois{y}{z}_0= 2 f x - d y + a z
\nonumber
\eea
This Lie-Poisson bracket ${\cal P}_0$ will play a relevant role in what follows, as it contains the information concerning the connection between equivalence classes (under group automorphisms) of the PL structures included in the generic bracket ${\cal P}{[a,b,c,d,e,f]}$ and the (tangent) Lie bialgebra structures on $\b$.

\subsection{Coboundary structures}

As a first step in the classification problem for ${\cal P}{[a,b,c,d,e,f]}$, we can investigate for which values of the parameters we have a coboundary PL group. Therefore, we write the most general skewsymmetric candidate for constant classical $r$-matrix on the book Lie  algebra $\b$:
\be
r=r^{12} (e_1 \wedge e_2)+ r^{13} (e_1 \wedge e_3)+r^{23} (e_2 \wedge e_3)
\label{rbook}
\ee
where $(r^{12},r^{13},r^{23})$ are free real parameters, and where we have to impose that $r$ is a solution of the mCYBE \eqref{mcybe}. We find that this equation gives no restriction on the $r^{ij}$  parameters, thus (\ref{rbook}) is the most general constant classical $r$-matrix for $\b$.

As a consequence, the Sklyanin bracket \eqref{sb} induced by  \eqref{rbook} can be computed by considering the left and right invariant vector fields on the book group 
\bea
L_1=L_y=e^{-x} \dfrac{\partial}{\partial y} \quad & L_2=L_z=e^{-x} \dfrac{\partial}{\partial z}\quad& L_3=L_x=\dfrac{\partial}{\partial x}\\
R_1=R_y=\dfrac{\partial}{\partial y}\quad& R_2=R_z=\dfrac{\partial}{\partial z}\quad&  R_3=R_x=\dfrac{\partial}{\partial x}- y  \dfrac{\partial}{\partial y} -z \dfrac{\partial}{\partial z}
\eea
and we finally obtain
\bea
&& \{x,y \}=r^{13} \left(1-\exp(-x)\right)
\cr
&& \{x,z \}=r^{23} \left(1-\exp(-x) \right)
\\
&& \{y,z \}=-r^{12} \left( 1-\exp(-2x) \right)-r^{23}y+r^{13}z .
\nonumber
\eea
After passing to the $X,Y,Z$ variables, we obtain the most general coboundary PL bracket
\bea
&& \{X,Y \}=r^{13}(X-1)X
\cr
&& \{X,Z \}=r^{23}(X-1)X
\\
&& \{Y,Z \}= r^{12} (X^2-1)- r^{23} Y+ r^{13}Z
\nonumber
\eea
which is a particular case of ${\cal P}{[a,b,c,d,e,f]}$ with 
\be
a=r^{13},
\quad b=0,
\quad c=0,
\quad d=r^{23},
\quad e=0,
\quad f=-r^{12}.
\ee
Therefore, we will have non-coboundary PL structures when either $b\neq 0$ or $c\neq 0$ or $e\neq 0$.

\subsection{Classification of Lie bialgebra structures on $\b$}

The essential tool in order to obtain the complete classification of PL structures on $C^\infty(G)$ is their  one-to-one correspondence with Lie bialgebra
structures on the Lie algebra $g=\mbox{Lie}(G)$. Thus, in order to get the full chart of equivalence classes of PL structures on G we have to look at the classification (under Lie algebra automorphisms) of Lie bialgebra structures on $\b$, which was fully completed in~\cite{gomez} for all real 3D Lie algebras.

If we look carefully at Gomez's classification for $\b$, by taking into account the following change of basis,
$$
\We_1=\Ge_1 \qquad \We_2=\Ge_2 \qquad \We_3=-\Ge_0.
$$ 
we get nine different classes of Lie bialgebra structures for $\b$, whose precise correspondence with our $(a,b,c,d,e,f)$ parameters is given in Table 1. For each of the Gomez cases, such correspondence is obtained by writing the cocommutator $\delta$ of the Lie bialgebra and by identifying its dual map $\delta^\ast$ with the linearized Poisson-Lie bracket \eqref{p0}. Thus, we conclude that we have nine inequivalent (under group automorphisms) classes of PL structures on the book group, that would be explicitly obtained by substituting the values of the $(a,b,c,d,e,f)$ parameters into the full PL bracket expressions \eqref{genericPL} or \eqref{pllocal}.

\begin{center}

    \begin{tabular}{ c | c | c | c | c | c | c | c |}
    & Cases in \cite{gomez} & $a$ & $b$ & $c$ & $d$ & $e$ & $f$ \\ 
     \hline
A   & $5$ ($\rho=1$)  & $0$ & $0$ & $0$ & $0$ & $0$ & $-1$ \\   
    \hline
B     & $6$ ($\rho=1$, $\chi=\Ge_0 \wedge \Ge_1$)   & $0$ & $0$ & $0$ & $-1$ & $0$ & $0$ \\   
    \hline
C     & $7$  ($\rho=1$) & $0$ & $\lambda$ & $0$ & $0$ & $0$ & $0$ \\   
    \hline
D     & $(1)$ & $0$ & $\lambda$ & $0$ & $0$ & $0$ & $-\alpha$ \\   
    \hline
E     & $(2)$ & $0$ & $0$ & $\lambda/2$ & $0$ & $\lambda/2$ & $-\omega$ \\
     \hline
F     & $9$ & $0$ & $0$ & $\lambda/2$ & $0$ & $\lambda/2$ & $0$ \\ 
    \hline 
G     & $10$ & $0$ & $0$ & $-1/2$ & $0$ & $0$ & $0$ \\ 
    \hline 
H     & $11$ & $0$ & $0$ & $-1/2$ & $0$ & $0$ & $-\omega$ \\ 
    \hline 
I     & $(3)$ & $0$ & $0$ & $-1/2$ & $-\alpha$ & $0$ & $0$ \\ 
    \hline 
    \end{tabular}

{\bf Table 1}. Correspondence with the classification~\cite{gomez} of Lie bialgebra structures on $\b$.
\end{center}

Several comments on Table 1 are in order:

\begin{itemize}

\item The parameters $\lambda,\alpha,\omega$ from~\cite{gomez} are nonzero real ones. As far as the equivalence classes are concerned, $\lambda$ is an essential parameter, $\alpha$ can be rescaled to any nonzero value and $\omega$ can be rescaled to any nonzero value of the same sign.

\item The only coboundary cases are $A$ and $B$: in fact, they correspond to the PL groups generated by the $r$-matrix with $f=r^{12}\neq 0$ and $d=r^{13}\neq 0$, respectively. 

\item Note that  the parameter $a$ always vanishes, since the Lie algebra automorphism that interchanges $\We_1$ and $\We_2$ would interchange $a$ and $d$. Therefore
the third coboundary case $a=r^{23}\neq 0$ is equivalent to case B.

\item The seven remaining cases are non-coboundary ones, and several of them are multiparametric.

\end{itemize}

For the sake of simplicity, in the rest of the paper we will use the $(a,b,c,d,e,f)$ parameters, although we will deal only with the nine inequivalent cases given in Table 1.

\subsection{The classical $r$-matrix approach}

It is interesting to note that, although the generic Poisson bracket ${\cal P}{[a,b,c,d,e,f]}$ is not a coboundary structure, it can be written in the classical $\hat r$-matrix form
\begin{equation}
\left\{ M \stackrel{\otimes}{,} M \right\}=\left[ M \otimes M , \hat r \right]
\end{equation}
with $\hat r$ being the $9 \times 9$ matrix
\begin{equation}
\hat r=\left(
\begin{array}{ccccccccc}
0 & -2 e & d & 2 e & 0 & 0 & -d & 0 & 0\\
c & b & a & 0 & e & 0 & 0 & -d & f\\
0 & 0 & 0 & 0 & 0 & e & 0 & -e & 0\\
-c & b & 0 & -2 b & -e & d & -a & 0 & -f\\
0 & -2 c & 0 & 2 c & 0 & a & 0 & -a & 0\\
0 & 0 &-c & 0 & 0 &-b & c & b & 0\\
0 & 0 & b & 0 & 0 & e & -b & -e & 0\\
0 & 0 & -c & 0 & 0 & 0 & c & 0 & 0\\
0 & 0 & 0 & 0 & 0 & 0 & 0 & 0 & 0
\end{array}
\right).
\end{equation}
As expected, the matrix $\hat r$ satisfies the classical Yang-Baxter equation only when $b=c=e=0$ (the coboundary cases). In that situation $\hat r$ is nilpotent of order $2$ and coincides with the $r$-matrix \eqref{rbook} taken in the fundamental representation \eqref{repbook}. Moreover,  the 
matrix $R=1+\hat r$ 
\begin{equation}
R=\left(
\begin{array}{ccccccccc}
1 & 0 & d & 0 & 0 & 0 & -d & 0 & 0\\
0 & 1 & a & 0 & 0 & 0 & 0 & -d & f\\
0 & 0 & 1 & 0 & 0 & 0 & 0 & 0 & 0\\
0 & 0 & 0 & 1 & 0 & d & -a & 0 & -f\\
0 & 0 & 0 & 0 & 1 & a & 0 & -a & 0\\
0 & 0 & 0 & 0 & 0 & 1 & 0 & 0 & 0\\
0 & 0 & 0 & 0 & 0 & 0 & 1 & 0 & 0\\
0 & 0 & 0 & 0 & 0 & 0 & 0 & 1 & 0\\
0 & 0 & 0 & 0 & 0 & 0 & 0 & 0 & 1
\end{array}
\right)
\label{quantumR}
\end{equation}
is an upper-triangular constant solution of the quantum Yang-Baxter equation $R_{12}R_{13}R_{23}=R_{23}R_{13}R_{12}$.

\section{A Lotka-Volterra PL group}

In this and the following sections we explicitly analyse the Poisson dynamics of three of the non-coboundary PL groups that we have identified in the previous section. In particular, let us now consider the case C in Table 1, namely ${\cal P}{[0,b,0,0,0,0]}$
given by the Poisson brackets
\be
\pois{X}{Y}= -b\, X Y
\qquad\qquad
\pois{X}{Z}= b\, X Z
\qquad\qquad
 \pois{Y}{Z}=b\, Y Z
\label{lvbook}
\ee
and the Casimir function
\be
\mathcal{C}=
 \dfrac{Y\, Z}
{X}.
\label{clv}
\ee

It is interesting to note that if we consider the new $\{J_3,J_\pm\}$ coordinates on the book group defined by
\be
X=e^{-2\pa\, J_3}
 \qquad\qquad
Y=e^{-\pa\, J_3}\,J_+
 \qquad\qquad
Z=e^{-\pa\, J_3}\,J_-
\label{coordst}
\ee
where $\pa\neq 0$ is an additional real parameter, the Poisson structure  \eqref{coordst} is rewritten as 
 \be
 \pois{J_3}{J_+}=\frac{b}{2\pa}\,J_+ 
 \qquad\qquad
 \pois{J_3}{J_-}=-\frac{b}{2\pa}\,J_-
  \qquad\qquad
 \pois{J_+}{J_-}=0,
 \label{poise2}
 \ee
the Casimir function reads $\mathcal{C}=J_+\,J_-$ and the coproduct map \eqref{bookcopr} is transformed into
  \be
\Delta(J_3)= 1 \otimes  J_3 +  J_3 \otimes 1
\qquad\qquad
\Delta( J_\pm)=e^{-\pa J_3} \otimes J_\pm + J_\pm \otimes e^{\pa J_3}
\label{coq}
\ee
where we recognize the usual shape of coproducts in 3D quantum algebras~\cite{CP,majid} provided that $\pa$ is interpreted as the `quantum' deformation parameter.

Moreover, \eqref{poise2} is just  the Poisson version of the (1+1) Poincar\'e-Hopf algebra in a `null-plane' basis (note that in the literature this is frequently called the pseudoeuclidean algebra, in which the parameter $b/2\pa$ can be reabsorbed through an automorphism).

\subsection{Integrable Lotka-Volterra dynamics}

An immediate application  arises from the fact that \eqref{lvbook} is a particular case of the three-parametric homogeneous quadratic Poisson structure~\cite{Nutku,ChrisDamianou, CDamianou}
\be
\begin{array}{lll}
\{X,Y\}=\alpha\, XY\qquad & \{X,Z\}= \beta\, X Z \qquad & \{Y,Z\}=\gamma\, Y Z
\end{array}
\nonumber
\ee
that underlies the integrability properties of a large family of 3D Lotka-Volterra systems (see~\cite{BBMLV} and references therein). In fact, 
if we consider the Hamiltonian function 
\be
\mathcal{H}=\alpha_{1}X+\alpha_{2}Y+\alpha_{3}Z+\beta_{1}\log X+\beta_{2}\log Y+\beta_{3}\log Z 
\label{hlv}
\ee
and the Poisson bracket \eqref{lvbook}, the Hamilton equations $
\dot{F}=\{F, \mathcal{H}\} 
$
read
\begin{eqnarray}
\dot{X}&=& b X\left[\alpha_{3}Z-\alpha_{2}Y  + (\beta_{3}-\beta_{2})\right]\notag\\
\dot{Y}&=& b Y\left[\alpha_{1}X+\alpha_{3}Z+(\beta_{1}+\beta_{3})\right]\label{lvsystem}\\
\dot{Z}&=&b Z\left[-\alpha_{1}X-\alpha_{2}Y-(\beta_{1}+\beta_{2})\right],
\notag
\end{eqnarray}
and this LV system has \eqref{hlv} and \eqref{clv} as integrals of the motion in involution.

Moreover, following~\cite{BBMLV} we can consider the full PL bracket on the book group ${\cal P}{[a,b,c,d,e,f]}$ as a {\em deformation} of ${\cal P}{[0,b,0,0,0,0]}$. In this way, the same Hamiltonian \eqref{hlv} leads to the five-parameter integrable perturbation of the LV system \eqref{lvsystem} given by:
\begin{eqnarray}
\dot{X}&=& b X \left[\alpha_{3}Z-\alpha_{2}Y+(\beta_{3}-\beta_{2})\right]
+\left(\alpha_{2}+\dfrac{\beta_{2}}{Y}\right)\left[
a X (X-1)-2c X Z 
\right]\cr
&& +\left(\alpha_{3}+\dfrac{\beta_{3}}{Z}\right)\left[d X (X-1)+2 e X Y \right]\notag\\
\dot{Y}&=&b Y \left[\alpha_{1}X+\alpha_{3}Z+(\beta_{1}+\beta_{3})\right]+a\left[
(\alpha_{3}Z+\beta_{3})-(X-1)(\alpha_{1}X+\beta_{1})
\right]\notag\\
&& \qquad +c Z\left[
2(\alpha_{1}X+\beta_{1})+(\alpha_{3}Z+\beta_{3})
\right]+\left(
\alpha_{3}+\dfrac{\beta_{3}}{Z}
\right)\left[Y (e Y-d)+f (1-X^{2})\right]\label{lvsystemdef}\\
\dot{Z}&=& b Z \left[ -\alpha_{1}X-\alpha_{2}Y- (\beta_{1}+\beta_{2})\right]+d\left[
(1-X)(\alpha_{1}X+\beta_{1})+(\alpha_{2}Y+\beta_{2})
\right]\notag\\
&& +\left(\alpha_{2}+\dfrac{\beta_{2}}{Y}\right)\left[
f (X^{2}-1)-Z (a + c Z)
\right]-e Y \left[2 (\alpha_{1}Y+\beta_{1})+(\alpha_{2}Y+\beta_{2})\right].
\notag
\end{eqnarray}
Again, both the Hamiltonian \eqref{hlv} and the Casimir function \eqref{casbook} are integrals of the motion in involution for this system.

Note that the system (\ref{lvsystemdef}) contains polynomial and rational perturbation terms with respect to (\ref{lvsystem}), and some choices of the `deformation' parameters $(a,c,d,e,f)$ could lead to regimes with saturation rates that could be worth studying from the integrable dynamics viewpoint (see~\cite{BF2} and references therein).

\subsection{Quantization}

Since the PL group \eqref{lvbook} is not a coboundary one, its quantization cannot be addressed through the quantum R-matrix formalism. Nevertheless a direct approach to the Hopf-algebra quantization of \eqref{lvbook} and \eqref{bookcopr} is feasible and leads to the $q$-commutation rules
\be
\hat{X}\hat{Y}-q^{-b}\hat{Y}\hat{X}=0
\qquad
\hat{X}\hat{Z}-q^{b}\hat{Z}\hat{X}=0
\qquad
\hat{Y}\hat{Z}-q^{b}\hat{Z}\hat{Y}=0,
\label{qcomm}
\ee
together with its compatible coproduct
\begin{eqnarray}
\Delta(\hat{X})&=&\hat{X} \otimes \hat{X}\notag\\
\Delta(\hat{Y})&=&\hat{X} \otimes \hat{Y} + \hat{Y} \otimes 1
\label{qcopr}\\
\Delta(\hat{Z})&=&\hat{X} \otimes \hat{Z} + \hat{Z} \otimes 1
\notag
\end{eqnarray}
where $q=e^{\pa}$. By assuming that $\hat{X}$ is invertible, the quantum Casimir operator for \eqref{qcomm} is found to be
\be
\hat{C}=\hat{X}^{-1}\hat{Y}\hat{Z}.
\label{qcas}
\ee
Obviously, if we define the `classical' limit of the commutation rules \eqref{qcomm} as
\be
\pois{a}{b}=\lim_{\pa\to 0}{\frac{\hat{a}\,\hat{b} - \hat{b}\,\hat{a}}{\pa}}
\ee
and by taking $\lim_{\pa\to 0}\hat{a}=a$ and $\lim_{\pa\to 0}\hat{b}=b$ we recover \eqref{lvbook}. Therefore, the non-commutative and non-cocom\-mutative Hopf algebra defined by \eqref{qcomm} and \eqref{qcopr} can be interpreted as a `quantum book group', since $\{\hat{X} ,\hat{Y} ,\hat{Z}\}$ are the non-commutative entries of the `quantum matrix'
\be
\hat{M}=\left(
\begin{array}{ccc}
 \hat{X} & 0 & \hat{Z} \\
 0 & \hat{X} & \hat{Y}\\
 0 & 0 & 1
\end{array} 
\right).
\label{qgroup}
\ee
In this context the `quantum book plane' would be the non-commutative algebra generated by $\{\hat{y} ,\hat{z}\}$ and such that
\be
\hat{y}\hat{z}-q^{b}\hat{z}\hat{y}=0,
\ee
since it can be straightforwardly checked that if we define the (left) co-action of $\hat{M}$ on the non-commutative space $\{\hat{y} ,\hat{z}\}$ in the form
\be
\left(
\begin{array}{c}
 \hat{y}' \\
  \hat{z}'\\
 1
\end{array} 
\right)
=
\left(
\begin{array}{ccc}
 \hat{X} & 0 & \hat{Z} \\
 0 & \hat{X} & \hat{Y}\\
 0 & 0 & 1
\end{array} 
\right)\otimes
\left(
\begin{array}{c}
 \hat{y}  \\
  \hat{z}\\
 1
\end{array} 
\right)
\ee
the covariance relation $\hat{y}'\hat{z}'-q^{b}\hat{z}'\hat{y}'=0$ holds.

Finally, we stress that by performing a nonlinear change of basis of the type \eqref{coordst} it is straightforward to prove that the Hopf algebra \eqref{qcomm} and \eqref{qcopr} is isomorphic to the quantum  (1+1) Poincar\'e algebra firstly introduced  in~\cite{VK,CGST} (see also~\cite{BCGST}) in which the commutation rules are just the (non-deformed) Poincar\'e ones analogous to \eqref{poise2} but the coproduct is deformed following \eqref{coq}.

\section{$sl_q(2)$ Poisson coalgebras as PL book groups}

In this section we will show that the Poisson analogues of the two $q$-deformations of $sl(2)$ are just the PL book groups D and I, provided certain specific values of the Lie bialgebra parameters are considered. In this way, the rich classical dynamics associated to these two $q$-Poisson deformations (see~\cite{BR,BHRplb,confseries} and references therein) can be interpreted in a Poisson-Lie context. Moreover, the quantization of the two PL groups will give rise to the two (standard and non-standard) quantum $sl(2)$ algebras.

\subsection{The standard $q$-deformation of $sl(2)$}

Let us start by applying the change of coordinates \eqref{coordst} to the full family of PL book groups \eqref{genericPL}. We obtain
 \bea
 && \pois{J_3}{J_+}=a\frac{\sinh(\pa J_3)}{\pa}+\frac{b}{2\pa}\,J_+ +\frac{c}{\pa}\,J_-
\nonumber \\[0.1cm]
 && \pois{J_3}{J_-}=d\frac{\sinh(\pa J_3)}{\pa}-\frac{e}{\pa}\,J_+ -\frac{b}{2\pa}\,J_-
\label{PLgenericpa} \\[0.1cm]
 &&\pois{J_+}{J_-}=2f\,{\sinh(2\pa J_3)}+\cosh(\pa J_3)
(-d\,J_+ + a\,J_-).
 \nonumber
 \eea
 where we recall that $\pa$ is an additional `deformation' parameter. In these new local coordinates the Casimir function \eqref{casbook} reads
\be
\mathcal{C}=2f\,\cosh(2\pa J_3)+2\sinh(\pa J_3)(-dJ_+ +  a J_- )
+ e\,J_+^2 + J_- (bJ_+ + c J_-)
\label{caspa}
\ee
and the coproduct map is \eqref{coq}, which is compatible with \eqref{PLgenericpa} for any value of the parameters.

Now, if we consider the non-coboundary D case ${\cal P}{[0,b,0,0,0,f]}$ and we choose
\be
b=2\pa 
\qquad\qquad
f=1/2\pa
\ee
we recover 
the `standard' $q$-deformation~\cite{DrICM,Ji,Ji2} of the $sl(2)$ Poisson algebra given by~\cite{BR,BCR}
 \be
\pois{J_3}{J_+}=J_+ 
 \qquad
 \pois{J_3}{J_-}= -J_-
  \qquad
 \pois{J_+}{J_-}=\frac{\sinh(2\pa J_3)}{\pa}
\label{qsl2st}
 \ee
where $q=e^\pa$ and the Casimir function \eqref{caspa} is shown to be proportional to the well-known $q$-deformed $sl(2)$ Casimir
\be
\tilde{\mathcal{C}}=\frac{\sinh^2(\pa J_3)}{\pa^2} + J_+\,J_- .
\label{qsl2cas}
\ee
Note that the limit $\pa\to 0$  of \eqref{qsl2st} leads to a linearized structure that is exactly the $sl(2)$ Lie-Poisson algebra. This is fully consistent with the book Lie bialgebra classification given in~\cite{gomez}, since its case $(1)$ gives $sl(2)$ as the dual Lie algebra arising from the cocommutator. Among the applications of the Poisson coalgebra \eqref{qsl2st} we recall the $q$-deformation constructed in~\cite{BR,BCR} of the Calogero-Gaudin system~\cite{Cal}.

Therefore, we can conclude that the Poisson analogue of the standard $q$-deformation of $sl(2)$ is just a specific PL structure on the book group, and the $q$-deformed coproduct \eqref{coq} is just the book group product law in a specific set $\{J_3,J_\pm\}$ of local coordinates. It is worth recalling that 
the  construction of Poisson $sl_q(2)$ as a PL group structure on $SB(2,C)$ was already given in~\cite{Marmo}, but here we obtain it as a particular case of the full book group classification of PL structures.


\subsection{The non-standard $q$-deformation of $sl(2)$}

The non-standard (or `jordanian') $q$-deformation of $sl(2)$ is generated by the twist operator firstly introduced in~\cite{Giaquinto1, Giaquinto2} (see also~\cite{Kulish} for the application of the Jordanian twist operator to integrable models). The Poisson analogue of this quantum algebra can be also obtained as the I case in the PL classification of Table 1 through the change of book group coordinates given by
\be
X=e^{-2\ph\, J_-}
 \qquad\qquad
Y=e^{-\ph\, J_-}\,J_+
 \qquad\qquad
Z=e^{-\ph\, J_-}\,J_3.
\label{nonst}
\ee
Now the generic PL book group reads
\bea
 && \pois{J_3}{J_+}=-2f\,{\sinh(2\ph J_-)}+\cosh(\ph\,J_-)(d\,J_+ -a\,J_3)
\nonumber \\[0.1cm]
 && \pois{J_3}{J_-}=-d\,\frac{\sinh(\ph J_-)}{\ph}+\frac{e}{\ph}\,J_++\frac{b}{2\ph}\,J_3
\label{PLns} \\[0.1cm]
 &&\pois{J_+}{J_-}= -a\,\frac{\sinh(\ph J_-)}{\ph}-\frac{b}{2 \ph}\,J_+-\frac{c}{\ph}\,J_3\nonumber
 \eea
with the coproduct
 \bea
\Delta(J_-)&=& 1 \otimes  J_- +  J_- \otimes 1\nonumber\\[0.1cm]
\Delta( J_{+})&=&e^{-\ph J_-} \otimes J_{+} + J_{+} \otimes e^{\ph J_-}
\label{cons}\\[0.1cm]
\Delta( J_{3})&=&e^{-\ph J_-} \otimes J_{3} + J_{3} \otimes e^{\ph J_-}
\nonumber
\eea
and the Casimir function
\be
\mathcal{C}=2f\,\cosh(2\ph J_-)+2\sinh(\ph J_-)(-dJ_+ + a J_3 )
+ e\,J_+^2 + J_3 (bJ_+ + c J_3).
\ee

Now, in order to identify the non-standard $q$-deformation ($q=e^\ph$) we take the non-coboundary I case ${\cal P}{[0,0,c,d,0,0]}$ and we choose
\be
c=-2\ph 
\qquad\qquad
d=1.
\ee
In this way we get the Poisson brackets
 \be
\pois{J_3}{J_+}=J_+ \,\cosh(\ph J_-)
 \qquad
 \pois{J_3}{J_-}= -\,\frac{\sinh(\ph J_-)}{\ph}
  \qquad
 \pois{J_+}{J_-}= 2\,J_3
 \label{nsbracket}
 \ee
whose Casimir can be shown to be proportional to the function
\be
\tilde{\mathcal{C}}=J_3^2 + J_+\,\frac{\sinh(\ph J_-)}{\ph}.
\ee
As expected, the limit $\ph\to 0$  of this deformation leads to a linearized structure that is again the $sl(2)$ Lie-Poisson algebra, since the corresponding case $(3)$ in~\cite{gomez} gives again $sl(2)$ as the dual Lie algebra. 

From the point of view of applications, we recall that the dynamics on ${\cal P}{[0,0,4 z,2,0,0]}$ has been extensively explored in~\cite{BHRplb}, where it has been shown that the $N$-dimensional Hamiltonian given by the $N$-th coproduct $\Delta^{(N)}$ of the $J_+$ coordinate
\be
H^{(N)}=\frac12 \Delta^{(N)}(J_+)
\ee
defines a (quasi-maximally) superintegrable geodesic flow on a ND hyperbolic space with non-constant scalar curvature $K$ given by
\be
K=-z\sinh^2(z\,r^2)
\ee
where $r$ is a radial coordinate of the space. Also, the geodesic flow on the ND hyperbolic space with constant scalar curvature $z$ can be obtained as another dynamical system defined on $N$ copies of ${\cal P}{[0,0,4 z,2,0,0]}$ and whose Hamiltonian is
\be
H^{(N)}=\frac12 \Delta^{(N)}(e^{z\,J_-}\,J_+).
\ee
Moreover, superintegrable potentials on these spaces have been obtained in~\cite{BHjpapot} by making use of the underlying coalgebra symmetry, that now can be interpreted as the invariance of the Poisson structure \eqref{nsbracket} under the book group multiplication.

\subsection{Quantization}

Similarly to the case of the Lotka-Volterra Poisson structure, the PL groups \eqref{qsl2st} and \eqref{nsbracket} are not coboundaries and cannot be quantized through the quantum $R$-matrix formalism. Nevertheless, the Hopf algebra quantization of both PL groups is well known: in the case of \eqref{qsl2st} is the standard (or Drinfel'd-Jimbo) quantum deformation of the universal enveloping algebra of sl(2)~\cite{DrICM,Ji,Ji2} and for \eqref{nsbracket} we obtain the non-standard or Jordanian quantum sl(2) algebra, whose $q$-deformed representation theory is also known~\cite{Aizawa}.

Moreover, the quantization of an algebra of the type \eqref{PLgenericpa} was obtained in~\cite{BCOjpa} by applying a power series Hopf algebra quantization procedure and by imposing order by order the compatibility between the deformed commutation rules and a coproduct map of the type \eqref{coq} (see eq.~(3.14) in~\cite{BCOjpa}). By taking the specific values of the parameters coming from Table 1, within this family of quantum algebras one can recover the quantum versions of the two-dimensional Euclidean and Poincar\'e algebras, as well as the quantum Heisenberg algebra (see~\cite{BCOjpa} and references therein). From the viewpoint of the present paper, all these three-dimensional quantum algebras are just Hopf algebra quantizations of different PL structures on the book group.


\section{The rest of PL structures on the book group}

A short disgression on the rest of the non-equivalent PL book groups is in order. Essentially, in all the cases Poisson analogues of the quantum deformations of many 3D real algebras are obtained. These algebras can be easily identified within the linearized PL structure which, in turn, is isomorphic to the dual lie algebra $g^\ast$ in~\cite{gomez}.

\subsection{The coboundary cases}

In the A case ${\cal P}{[0,0,0,0,0,f]}$  the PL group would be given by 
 \be
\{X,Y\}=0 
\qquad
 \{X,Z\}=0
\qquad
\{Y,Z\}= f (1-X^2)
 \label{PLA}
 \ee
with Casimir 
$
\mathcal{C}=
f(1+X^{2})/
{X}
$, which means that $X$ is also a central function. In the coordinates \eqref{coordst} we have
 \be
\pois{J_3}{J_+}=0
\qquad
\pois{J_3}{J_-}=0
\qquad
\pois{J_+}{J_-}=2f\,{\sinh(2\pa J_3)}
\label{plheis}
 \ee
and if we take $f=1/4\pa$ together with \eqref{coq} we obtain the Poisson analogue of the quantum Heisenberg algebra firstly introduced in~\cite{CGSTheis}. This coboundary Poisson structure was already presented in~\cite{BHOSpl}, and has been also recently considered in~\cite{checos}.

A similar analysis can be performed for the B structure  ${\cal P}{[0,0,0,d,0,0]}$  (we recall that the case ${\cal P}{[a,0,0,0,0,0]}$ is equivalent to B through an automorphism). We get
 \be
\{X,Y\}=0 
\qquad
 \{X,Z\}=d X\,(X - 1)
\qquad
\{Y,Z\}= -d Y.
 \label{PLB}
 \ee
with Casimir 
$
\mathcal{C}=
d\,Y\,(X-1)/
{X}
$. In the coordinates \eqref{nonst} we have
\be
\pois{J_3}{J_+}=d\,J_+\,\cosh(\ph\,J_-)
\qquad
\pois{J_3}{J_-}=-d\,\frac{\sinh(\ph J_-)}{\ph}
\qquad
\pois{J_+}{J_-}= 0,
\label{ple2ns}
 \ee
together with \eqref{cons}, and if we take $d=1$ we get the Poisson analogue of the non-standard quantum deformation of the (pseudo)Euclidean algebra given in~\cite{BHPOS}. 

Since both PL groups are coboundaries, the Hopf algebra quantization problem for \eqref{PLA} and \eqref{PLB} can be directly addressed through the RTT formalism~\cite{FRT} by making use of the quantum $R$-matrix \eqref{quantumR}. Moreover, the star-product quantization of \eqref{plheis} was presented in~\cite{BHOSpl}, and the quantum analogue of \eqref{ple2ns} is also known (see~\cite{BCOjpa,BHPOS} and references therein).

 
 \subsection{The non-coboundary ones}
 
 The case E corresponds to  ${\cal P}{[0,0,c,0,c,f]}$ and the PL bracket coming from \eqref{coordst} is
 \be
\pois{J_3}{J_+}=\frac{c}{\pa}\,J_-
\qquad
\pois{J_3}{J_-}=-\frac{c}{\pa}\,J_+ 
\qquad
\pois{J_+}{J_-}=2f\,{\sinh(2\pa J_3)}
 \label{PLE}
 \ee
where the Casimir function is
$
\mathcal{C}=2f\,\cosh(2\pa J_3)
+ c\,J_+^2 + c J_-^2
$. By making $c=\pa$, $f$ proportional to $1/\pa$  and depending on the sign of $f$ we get the Poisson analogue of the (standard) quantum $so(3)$ or $so(2,1)$ algebras (see~\cite{BHOS} and references therein).

Case F is  ${\cal P}{[0,0,c,0,c,0]}$, which is just the $f\to 0$ limit of the previous PL group \eqref{PLE}, that correspond to a Poisson euclidean algebra with non-deformed commutation rules (albeit with deformed coproduct given by \eqref{coq}). 

Case G is obtained by considering ${\cal P}{[0,0,c,0,0,0]}$, and from \eqref{coordst} we get
 \be
\pois{J_3}{J_+}=\frac{c}{\pa}\,J_-
\qquad
\pois{J_3}{J_-}=0 
\qquad
\pois{J_+}{J_-}=0
 \label{PLG}
 \ee
which is a (non-deformed) Poisson-Heisenberg algebra with deformed coproduct \eqref{coq}.

Finally, case H is the PL group given by ${\cal P}{[0,0,c,0,0,f]}$
 \be
 \pois{J_3}{J_+}=\frac{c}{\pa}\,J_-
\qquad
\pois{J_3}{J_-}=0
\qquad
\pois{J_+}{J_-}=2f\,{\sinh(2\pa J_3)} 
\label{plH}
\ee 
with Casimir 
$
\mathcal{C}=2f\,\cosh(2\pa J_3) + c\,J_-^2
$. Under analogous assumptions for the $c$ and $f$ parameters we get Poisson analogues of the standard quantum deformations of the euclidean $e(2)$ and pseudo-euclidean $e(1,1)$ algebras given in~\cite{BHOS}.


\section{Concluding remarks}

We have presented a systematic construction and classification of the PL structures on the real 3D book group. Since the book Lie algebra is not semisimple, we have found that most of such PL groups are non-coboundary ones. Nevertheless, we have shown that these non-coboundary PL structures do present interesting dynamical features and can be quantized as Hopf algebras, which motivates a more detailed study of PL structures on other non-simple Lie groups.

Namely, the full classification and construction of PL structures on all real 3D Lie groups will be presented elsewhere~\cite{BBM3d}, and the dominant role played in the non-semisimple cases by non-coboundary structures will arise again, as expected from~\cite{gomez}. In this context, it is important to stress that working on different groups would imply to consider different comultiplication maps $\Delta$ under which the corresponding PL structures will be invariant.

This classification program of PL groups is also motivated by the physical significance of quadratic (Poisson) algebras in the theory of integrable and/or exactly solvable Hamiltonian systems (see, for instance, ~\cite{Maillet,Fokas, Dask}). As we have shown in the case of Lotka-Volterra equations, it could happen that many relevant quadratic Poisson algebras could find a group theoretical interpretation as PL structures on certain (possibly non-simple) Lie groups.

Finally, it is interesting to mention that the  book Lie algebra $\b$ is isomorphic to the so-called (2+1) $\kappa$--Minkowski non-commutative spacetime~\cite{kpoin1, kpoin2} 
\be
\conm{{\hat x}_1}{{\hat x}_0}=\frac{1}{\kappa}\,{\hat x}_1
\qquad
\conm{{\hat x}_2}{{\hat x}_0}=\frac{1}{\kappa}\,{\hat x}_2
\qquad
\conm{{\hat x}_1}{{\hat x}_2}=0
\ee
where ${\hat x}_0$ is the (non-commutative) time `coordinate', ${\hat x}_1$ and ${\hat x}_2$ are the space ones and $\kappa$ is the quantum deformation parameter related with the Planck energy scale. In this context it has been noticed in~\cite{Freidel} that the ordered plane waves on $\kappa$-Minkowski spacetime would be given by $e^{i\,k^\mu \,{\hat x}_\mu}$, which can be properly regarded as a book-group element $M$ \eqref{group} and where our coordinates $(z,y,x)$ would be directly related with the components of the wave vector. Therefore, the possible dynamical significance of all the PL structures here presented is worth to be studied from this perspective, which is also motivated by the recent interest on the role of PL structures and Lie bialgebras in 3D gravity~\cite{MSjmp,AMII, FR,cm2, BHMplb}.


\section*{Acknowledgements}

This work was partially supported by the Spanish MICINN   under grant   MTM2010-18556
and by INFN--MICINN (grant AIC-D-2011-0711).    



\end{document}